\documentclass[letterpaper, 12pt]{ewshm}
\pdfoutput=1
\usepackage{scalerel}
\usepackage{graphicx}
\usepackage{xcolor}
\usepackage{amsfonts}
\usepackage{amsmath}
\usepackage{amssymb}
\usepackage{times}
\usepackage{cite}
\usepackage{hhline}
\usepackage{compactbib}
\usepackage{subcaption}
\usepackage{booktabs, tabularx}
\newcolumntype{C}{>{\centering\arraybackslash}X}
\usepackage[utf8]{inputenc}
\usepackage{dirtytalk}
\usepackage{comment}

\AtEndDocument{\par\leavevmode}

\renewcommand{\thefigure}{\arabic{figure}}
\renewcommand{\thetable}{\Roman{table}}

\long\def\symbolfootnote[#1]#2{\begingroup%
\def\thefootnote{\fnsymbol{footnote}}\footnote[#1]{#2}\endgroup}

\makeatletter
\renewcommand\@biblabel[1]{#1.}
\makeatother

\begin{document}

 \vspace{\baselineskip}
 \vspace{\baselineskip}
\begin{center}
	\Large{\textbf{Nonlinear Reduced Order Modelling of Soil Structure Interaction Effects via LSTM and Autoencoder Neural Networks
}}\\
	\normalsize
	\vspace{\baselineskip}
	\vspace{\baselineskip}
	\vspace{\baselineskip}
	\vspace{\baselineskip}
	Thomas Simpson$^{1}$, Nikolaos Dervilis$^{2}$, Philippe Couturier$^{3}$, Nico Maljaars $^{4}$ and Eleni Chatzi$^{1}$\\ 
	\vspace{\baselineskip}
	$^{1}$ Institute of Structural Engineering, Department of  Civil, Environmental and Geomatic Engineering, ETH Zürich, Stefano-Franscini Platz 5, 8093, Zürich, Switzerland\\
	$^{2}$ Dynamics Research Group, Department of Mechanical Engineering, University of Sheffield,  Mappin Street, Sheffield S1 3JD, England\\
	$^{3}$ Siemens Gamesa Renewable Energy, USA\\
	$^{4}$ Siemens Gamesa Renewable Energy, Netherlands
\end{center}
\normalsize
\vspace{\baselineskip}
\vspace{\baselineskip}
\vspace{\baselineskip}

\section*{ABSTRACT}
\label{abstract}

In the field of structural health monitoring (SHM), inverse problems which require repeated analyses are common. With  the increase in the use of nonlinear models, the development of nonlinear reduced order modelling techniques is of paramount interest. Of considerable research interest, is the use of flexible and scalable machine learning methods which can learn to approximate the behaviour of nonlinear dynamic systems using input and output data. One such nonlinear system of interest, in the context of wind turbine structures, is the soil structure interaction (SSI) problem. Soil demonstrates strongly nonlinear behaviour with regards to its restoring force and has been shown to considerably influence the dynamic response of wind turbine structures. In this work, we demonstrate the application of a recently developed nonlinear reduced order modelling method, which leverages Autoencoder and LSTM neural networks, to a nonlinear soil structure interaction  problem of a wind turbine monopile subject to realistic loading at the seabed level. The accuracy and efficiency of the methodology is compared to full order simulations carried out using Abaqus. The ROM was shown to have good fidelity and a considerable reduction in computational time for the system considered.


\vspace{12pt} 
\newpage
\noindent \uppercase{\textbf{Introduction}}  \vspace{12pt} 

The inverse formulations required within the context of structural health monitoring require repeated and rapid analyses of dynamic models. With the increased use of nonlinear models, the computational burden of these repeated analyses can become prohibitive. This motivates the use of reduced order modelling (ROM) methods, which can approximate these forward simulations with sufficient accuracy, whilst reducing the computational burden. There has been considerable research interest in the leveraging of machine learning (ML) methods for delivery of efficient ROMs \cite{LAGAROS201292,SWISCHUK2019704}. This involves training ML models from full order simulations and using the powerful capacity of ML to capture nonlinear relationships in data, such that these models can then be used to approximate the response of the full order model (FOM) under different inputs \cite{vega,Simpson}.

One such nonlinear problem, in the context of wind turbine structures, is the soil structure interaction (SSI) problem. The effect of soil stiffness has been shown to bear significant effects on the dynamic response of wind turbines, which is not always beneficial \cite{HARTE2012509}, as was often previously assumed. A common method to computationally represent SSI is the py curve method. The soil resistance is represented by multiple nonlinear springs, which restrain the motion of the foundation. The properties of the nonlinear springs vary with depth and are defined by relations that are experimentally derived.

The majority of existing ROM techniques involve projection onto linear bases with the most widely used methods being the Craig-Bampton and MacNeal-Rubin methods \cite{Craig1968,DeKlerk2008}. These methods make use of linear, or linearised, modes as reduction bases and are widely implemented in finite element software \cite{Ansys}. Such methods are well established for linear systems, but when considering nonlinear systems their application becomes more difficult. Targeted schemes for nonlinear systems are available, which involve the use of multiple linear reduction bases \cite{VLACHAS2021116055,Amsallem} or the enrichment of traditional bases \cite{Wu2016}. However, these methods tend to be limited to the types of nonlinearity to which they are applicable and require extensive knowledge of the modelled system.

In this work, we demonstrate exploitation of a recently developed nonlinear ROM framework \cite{Simpson}. The developed framework makes use of autoencoder neural networks for dimensionality reduction \cite{Hinton2006}; LSTM neural networks \cite{Hochreiter1997} are then used to model the dynamics of the system on the reduced basis. The ROM method is based purely on input and output data from training simulations carried out with the FOM. This then allows for the system response to be approximated for new forcing input time histories. 

We demonstrate the ROM framework for the problem of nonlinear SSI for a realistic wind turbine monopile. The exemplar system consists of linear beam elements representing the monopile; at each node, the monopile is constrained laterally with nonlinear spring elements defined by the soil stiffness. The monopile is excited by 2 lateral forces and 2 moments applied at the seabed level, which corresponds to the top structural node of the section of the monopile modeled. We demonstrate how a time series ROM of the system can be constructed on a reduced basis extracted by the autoencoder. The accuracy and efficiency of the ROM is compared to the full order simulation carried out using Abaqus Finite Element software \cite{Abaqus}. 

The paper consists of the following sections, first the soil structure interaction problem considered is presented and the finite element modelling explained. Following this, the ROM methodology and its application to this problem is explained. The next section details the procedure for training of the ROM. We then compare the results of the ROM to the FOM on a testing dataset and compare the computational time. The final section gives some concluding remarks and future work.

\vspace{12pt}
\noindent \uppercase{\textbf{Soil Structure Interaction Problem}}\vspace{12pt}

The SSI problem is of significant interest as it exhibits considerable nonlinearity in the response and can have significant impact on the response of a turbine. Due to its inherently nonlinear nature, the modeling of such SSI problems requires a nonlinear analysis which are more computationally expensive. As such, the SSI problem can prove to be a bottleneck in coupled analyses of wind turbine systems. Whilst the rest of the structure can be analysed more rapidly, the overall computational time is dictated by the nonlinear analysis of the SSI. 

\begin{figure}[h!]
    \centering
    \begin{subfigure}[t]{0.25\textwidth}
        \centering
        \includegraphics[width=11.5mm]{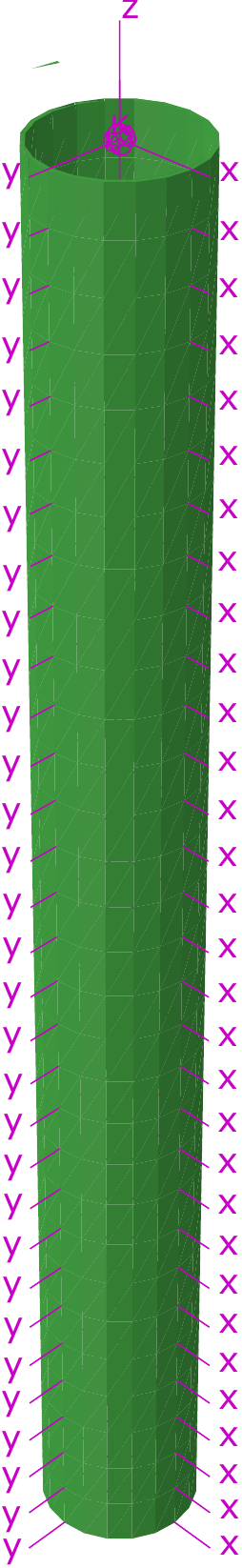}
        \caption{}
        \label{fig:femono}
    \end{subfigure}%
    ~ 
    \begin{subfigure}[t]{0.75\textwidth}
        \centering
        \includegraphics[width=\textwidth]{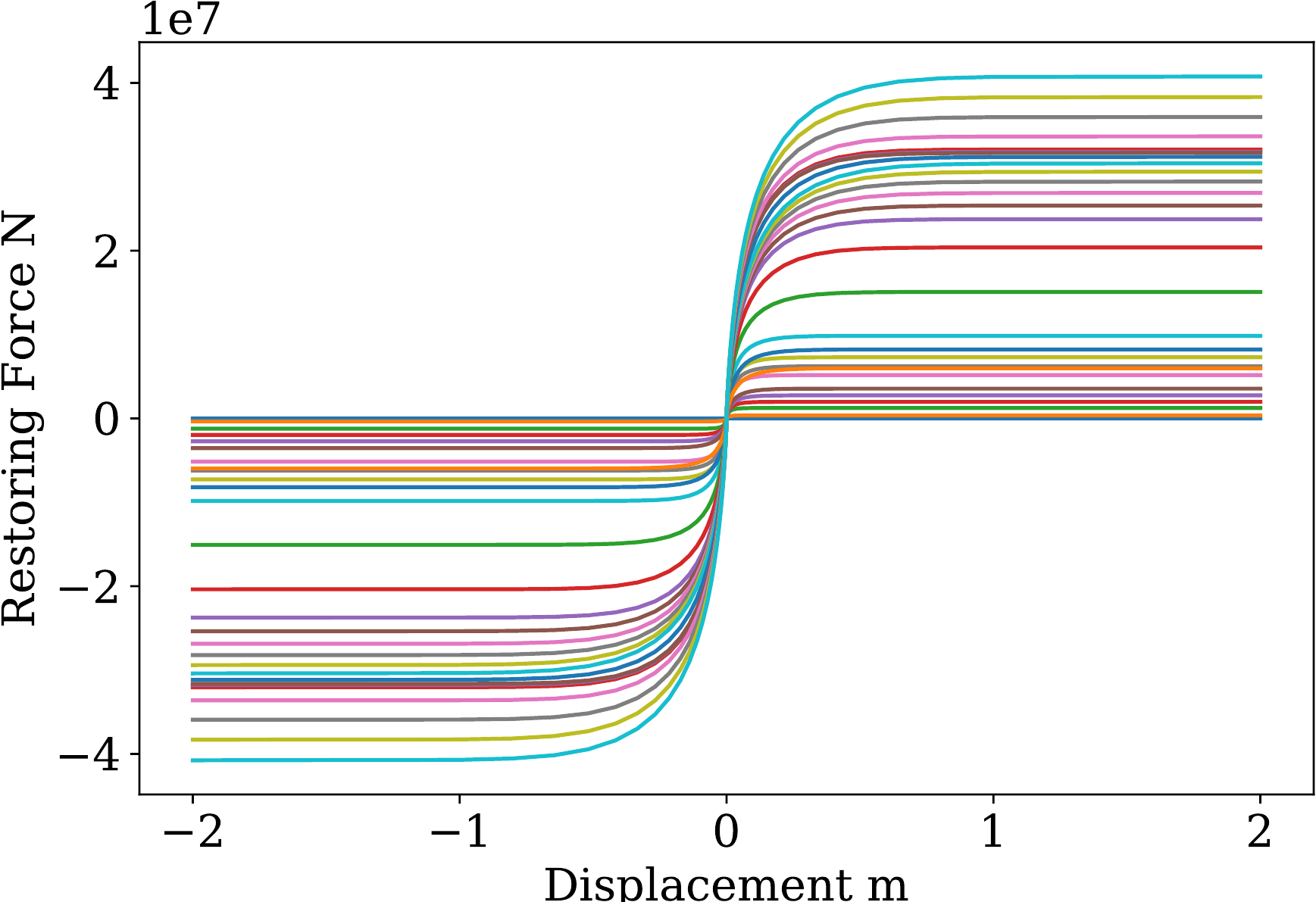}
        \caption{}
        \label{fig:soilspring}

    \end{subfigure}
    \caption{(a) The finite element model of the steel monopile. (b) The nonlinear restoring force of the soil at various depths. }
    \label{fig:ROMss}
\end{figure}

The SSI problem examined herein consists of a wind turbine monopile constrained by soil which exhibits a nonlinear stiffness. The monopile's boundary condition is assumed at \say{ground level}, with forcing applied at this \say{ground level} node. By isolating the monopile/soil system for the analysis, the remains of the wind turbine structure is represented only as this exogenous forcing. The monopile itself has a length of 30 m, diameter of 9.5 m and wall thickness of 0.08 m. It is considered to be constructed of steel and is modelled using 30 linear beam elements, of 1 m length, as shown in Figure \ref{fig:femono}. The soil is modelled using nonlinear springs which constrain the motion of the monopile in both the x and y axis. The restoring forces of these springs are defined according to nonlinear relations provided by Siemens Gamesa and are depth dependent. The restoring forces of the springs at each node are plotted in Figure \ref{fig:soilspring}. The exogenous forcing applied at the \say{ground level} node is applied as forcing along both the x and y directions and bending moment about the x and y axis giving a total of 4 input vectors.

\begin{figure}[h!]
    \centering
    \includegraphics[trim={0mm 0mm 0mm 2mm
    },clip,width=0.75\textwidth]{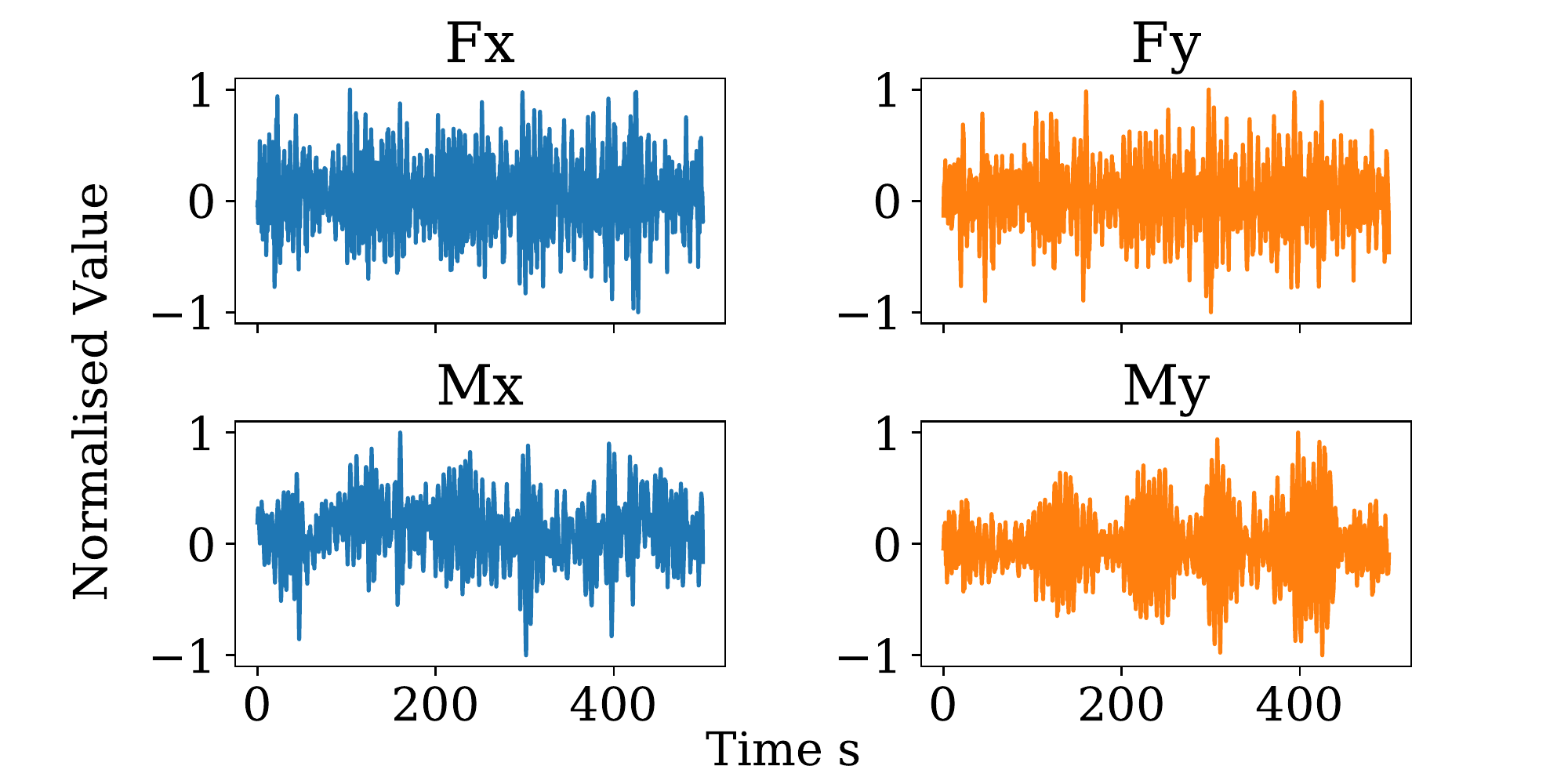}
    \caption{Example snapshots of the input forcing and bending moment time histories used in the analysis.}
    \label{fig:my_label}
\end{figure}

\vspace{12pt} 
\noindent \uppercase{\textbf{Reduced Order Modelling Methodology}} \vspace{12pt} 

\begin{figure}[h!]
    \centering
    \includegraphics[width=.9\textwidth]{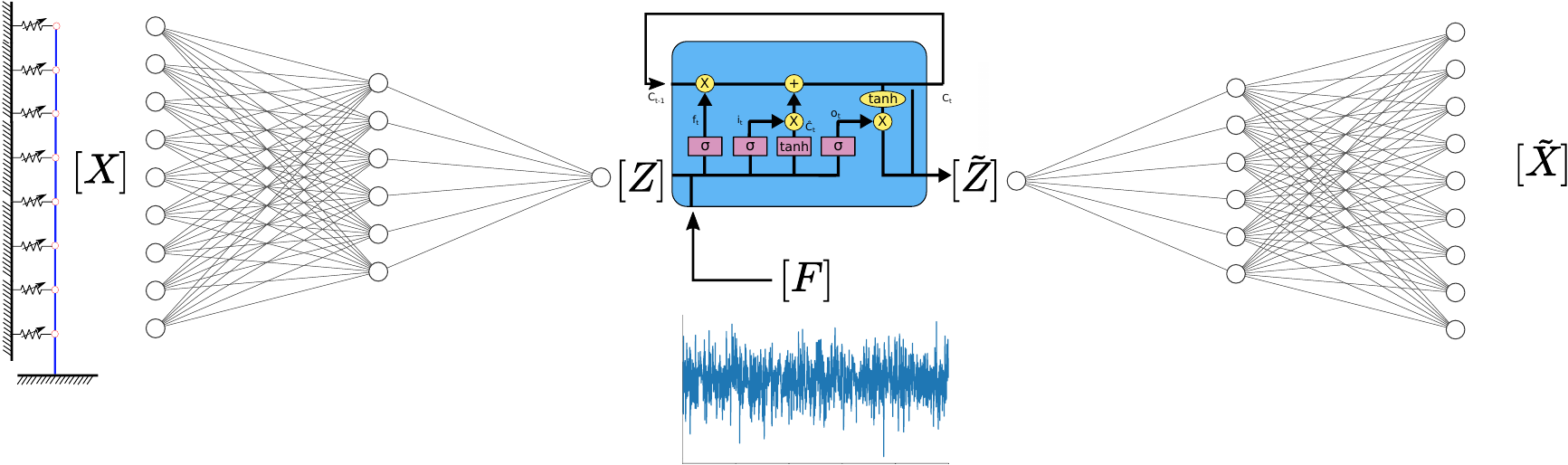}
    \caption{AE-LSTM based ROM framework in training mode.}
    \label{fig:ROM_Method}
\end{figure}

The reduced order modelling method used herein is a purely data-driven approach, in the sense that it relies on response time histories of input and output data, which in this case stem from FOM simulations, but that would further be possible to retrieve via measurements. The method makes use of an autoencoder neural network, for the purpose of dimensionality reduction and an LSTM neural network for the learning of system dynamics \cite{Simpson}. The process is demonstrated in figure \ref{fig:ROM_Method}, where the training of the ROM initially involves the generation of input-output data using the FOM of the system of concern. The first step of training the ROM involves taking the output data, in this case the $x$ and $y$ displacement time histories at each of the 31 nodes of the FE model, this data is represented as the matrix $[X]$. This output-only data is then used to train an autoencoder neural network to reduce the dimensionality of the original space. An autoencoder neural network is a neural network variant that is often used for dimensionality reduction or denoising problems \cite{Hinton2006}. 
The autoencoder network learns a nonlinear transform, which reduces the 62 DOF displacement data to a reduced number, $n$, of latent variables represented as $[Z]$. The value of $n$ is considerably smaller than 62 and is to be chosen as a hyperparameter. Simultaneously, the autoencoder learns the inverse operation; this is a nonlinear transform which returns from the reduced latent space $[Z]$ to the physical DOFs $[X]$.

\begin{figure}[h!]
    \centering
    \includegraphics[width=0.65\textwidth]{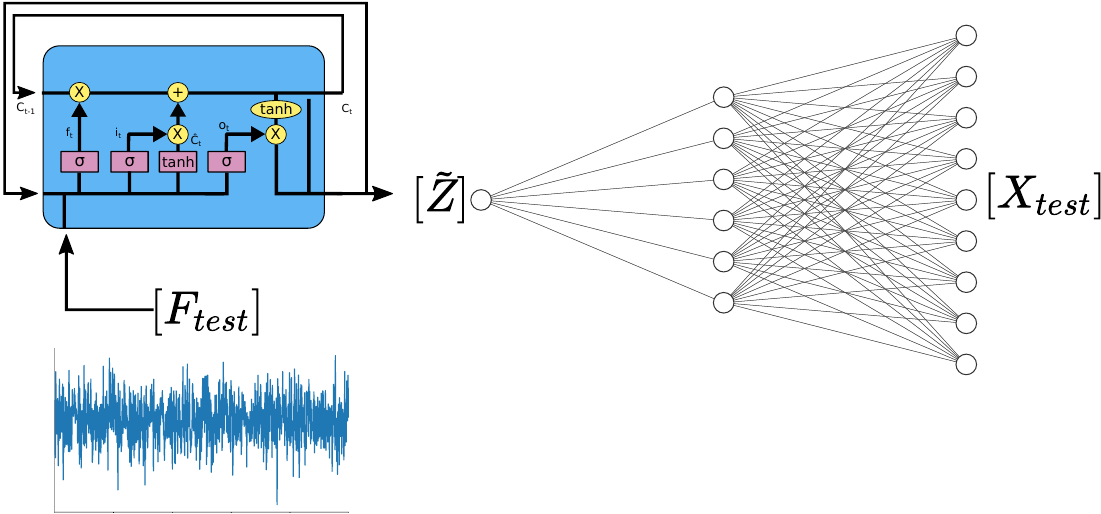}
    \caption{AE-LSTM based ROM framework in testing mode.}
    \label{fig:ROMtest}
\end{figure}

Following the training of the Autoencoder, an LSTM network is trained to learn the dynamics of the system. LSTM neural networks are a variant of a recurrent neural network, specifically designed for making predictions on sequence data and learning long term effects \cite{Hochreiter1997}. Concretely, the LSTM network learns to predict the response of the system, within the reduced latent space $[Z]$, taking as exogenous input the forcing time series collected in matrix $[F]$. The key here is that the LSTM network must only predict response within the reduced coordinate space which results in a much more lightweight and trainable model. Finally, in order to use the ROM for prediction, as illustrated in Figure \ref{fig:ROMtest}, a new set of forcing time series are given to the LSTM network. The network then predicts response of the system within the latent space, $[\Tilde{Z}]$, and the predicted response of the full system is recovered by passing $[\Tilde{Z}]$ through the decoder to recover the prediction in the physical space $[\Tilde{X}]$. The full methodology is extensively described and demonstrated in \cite{Simpson}.

\vspace{12pt} 
\noindent \uppercase{\textbf{ROM Training}} \vspace{12pt} 

For the purpose of training and testing the ROM, 12 sets of forcing and bending moment time series were provided from Siemens Gamesa. These forces and moments, which are considered as inputs to the monopile structure at the ground level node, are taken from coupled simulations between an aeroelastic model and an existing foundation model. The different datasets are taken from simulations performed at different amplitude of wind and wave loading: from normal to extreme conditions, and different operating conditions including normal production and idle. For each of the 12 time series, dynamic finite element simulations were carried out as previously specified, in order to create 12 sets of input output data. 11 of these datasets were used for training the ROM and 1 set for testing the ROM.

Firstly, the Autoencoder network was trained to extract a low-dimensional representation of the response data. When constructing the autoencoder, a key hyperparameter to select is the number of neurons in the bottleneck layer. This defines the number of latent variables in the reduced representation of the system. This value is selected as a balance of reduction and fidelity achieved by the ROM. However, the size and number of hidden layers in the autoencoder are also of vital importance to the performance, along with the activation functions used. In this case, it was chosen to retain 4 latent variables in the compressed representation as this was found to adequately recreate the important response of the system. Having chosen this bottleneck layer size, the network architecture was then optimised. The final autoencoder used, consisted of five layers: a 62 dimensional input layer with linear activation functions, a 62 dimensional hidden layer with \textit{\textit{tanh}} activation functions, a 4 dimensional bottleneck layer with \textit{tanh} activation functions, a 62 dimensional hidden layer with \textit{\textit{tanh}} activation functions followed by a 62 dimensional output layer with linear activation functions.The network was trained using the ADAM optimisation algorithm \cite{kingma2014adam}.

An LSTM network was subsequently trained, which learns to predict the response of the system in the latent space given the the forcing inputs. For the regression, a single \textit{tanh}-activated LSTM unit was used, with a cell state size of 45, with a 4 dimensional fully connected layer as an output layer with linear activation function. The LSTM network was herein set up in an explicitly auto-regressive manner, which is not typically the case. As such, in order to make a prediction at each time step the network input was not only fed the exogenous (forcing) variables and the hidden state variable from the previous time step, but was further explicitly passed the output (NNM response) values from the previous time steps. In training mode, these previous values were passed in prediction error manner, meaning that the exact output values of the system at time $t_{k-1}$ were passed on as inputs at time $t_k$. On the testing datasets, however, a simulation error type arrangement was used meaning that the \emph{predicted} outputs at time $t_{k-1}$ were passed as inputs for prediction at time $t_k$. During the prediction phase, the input was formulated again after each prediction to include the autoregressive features, with repeated one step ahead predictions. When training an RNN using the back propagation through time (BPTT) algorithm, a maximum number of time steps to back propagate error for each training example must be given. The number of time steps considered is another hyperparameter of the network to be chosen for optimum performance. In this case, 100 time steps were included for the BPTT algorithm. 

\vspace{12pt}
\noindent \uppercase{\textbf{Full Rom Performance}} \vspace{12pt} 

\begin{figure}[t!]
    \centering
    \begin{subfigure}[t]{0.5\textwidth}
        \centering
        \includegraphics[trim={9mm 3mm 20mm 5mm},clip,width=0.95\textwidth]{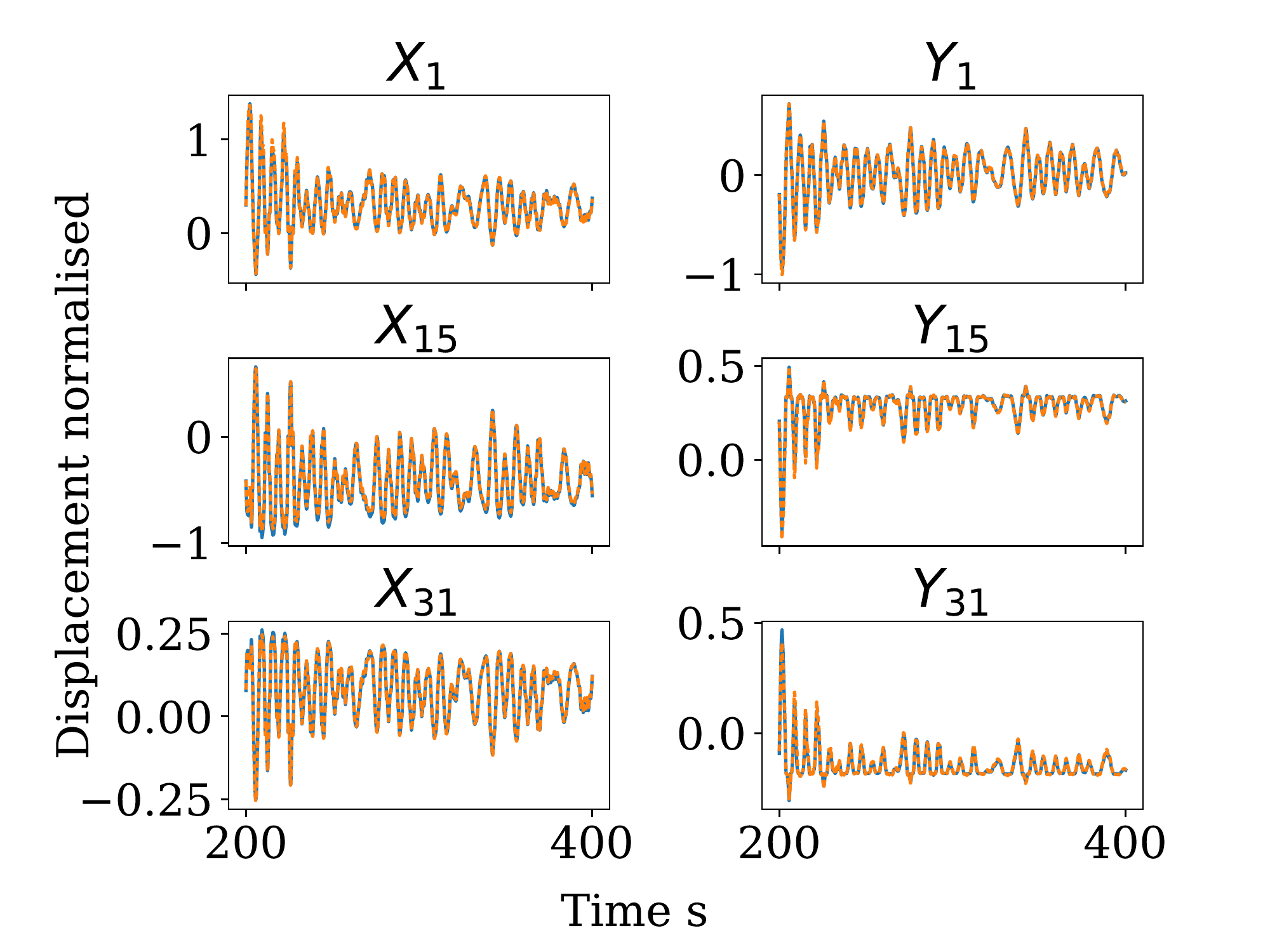}
        \caption{Steady-state response}
        \label{fig:autoencoderSS}
    \end{subfigure}%
    ~ 
    \begin{subfigure}[t]{0.5\textwidth}
        \centering
        \includegraphics[trim={9mm 3mm 20mm 5mm},clip,width=0.95\textwidth]{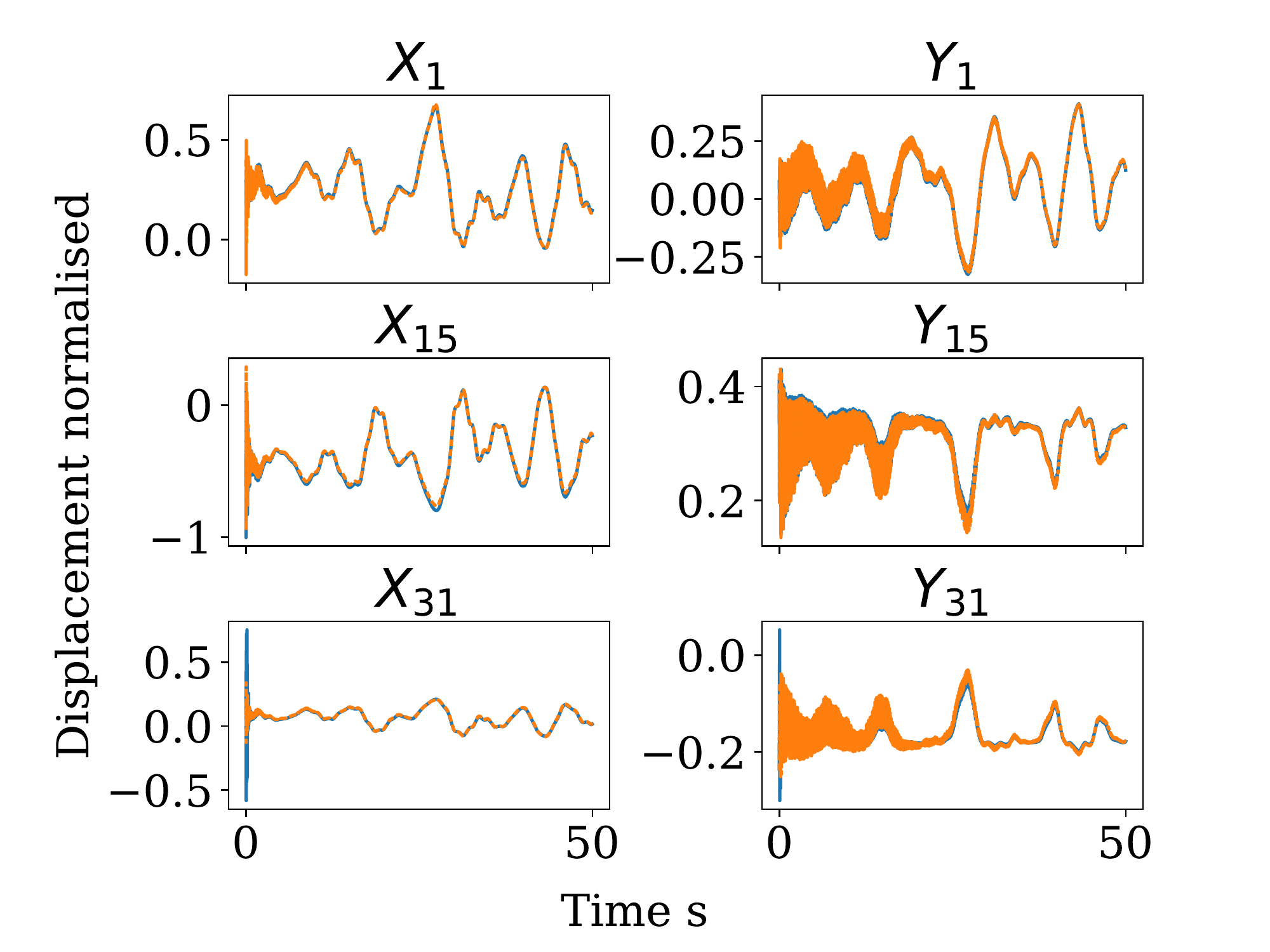}
        \caption{Initial Transient Response}
        \label{fig:autoencoderTransient}
    \end{subfigure}
    \caption{Original displacement (blue line) vs. displacement reconstructed by an autoencoder comprising 4 nodes in the bottleneck layer (orange dashed line).}
    \label{fig:autoencoder}
\end{figure}

The fidelity of the autoencoder is an upper limit on the accuracy of the ROM, Figure \ref{fig:autoencoder} compares full order response, to the response reconstructed using an autoencoder with 4 units in the hidden layer. The DOFs presented comprise of the $x$ and $y$ response at the mud line for $X_1$ and $Y_1$, at a depth of 15 m for $X_{15}$ and $Y_{15}$ and a depth of 30 m for $X_{30}$ and $Y_{30}$. The fidelity of the response can be seen, in Figure \ref{fig:autoencoderSS}, to be very good in the steady state. In the initial transient regime after the sudden application of forcing, however, it can be seen, from Figure \ref{fig:autoencoderTransient}, that the autoencoder fails to capture the high frequency transient response.

\begin{figure}[t!]
    \centering
    \begin{subfigure}[t]{0.5\textwidth}
        \centering
        \includegraphics[trim={9mm 3mm 20mm 5mm},clip,width=0.95\textwidth]{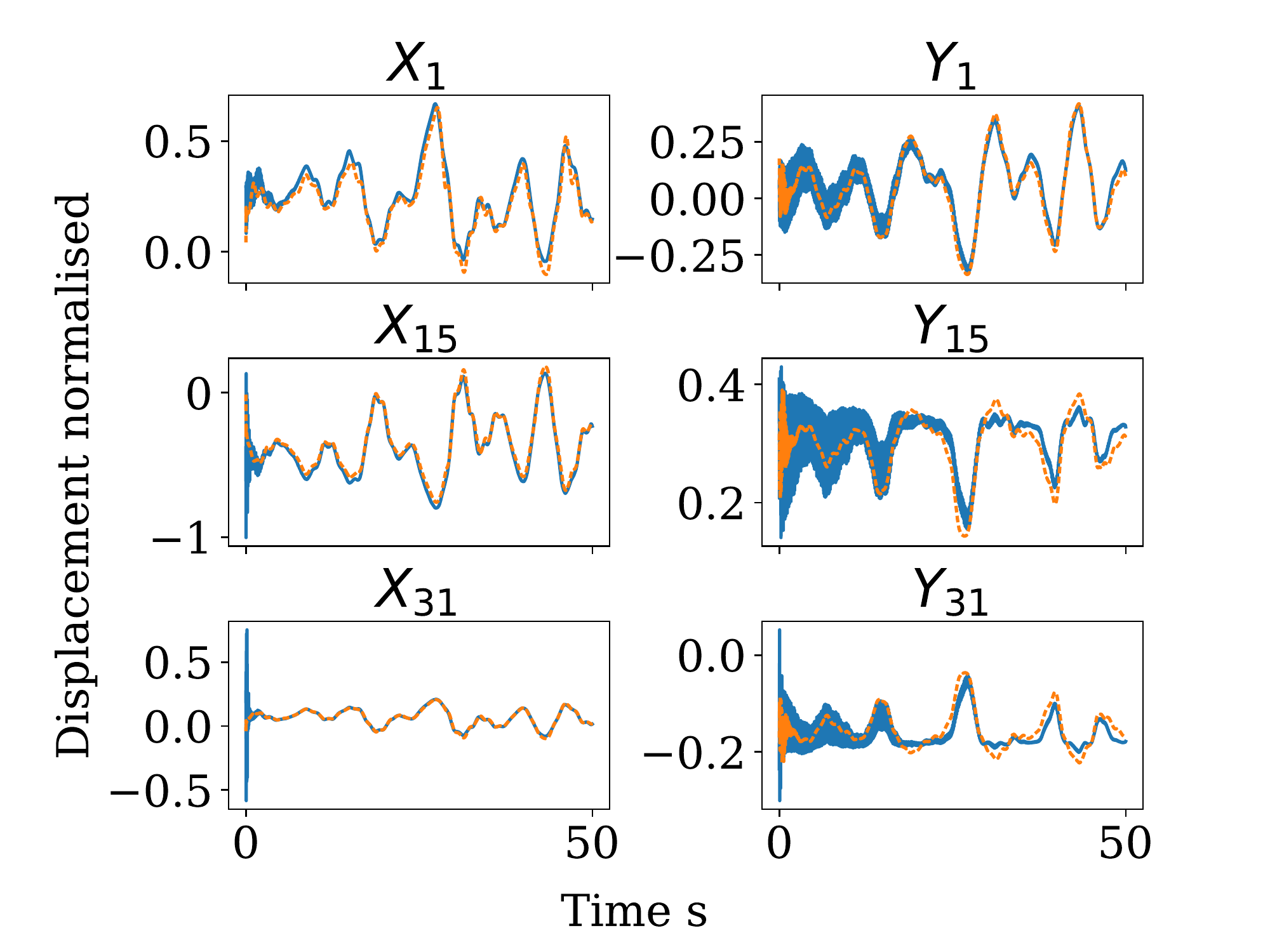}
        \caption{Time series response}
        \label{fig:ROMtestTransient}
    \end{subfigure}%
    ~ 
    \begin{subfigure}[t]{0.5\textwidth}
        \centering
        \includegraphics[trim={9mm 3mm 20mm 5mm},clip,width=0.95\textwidth]{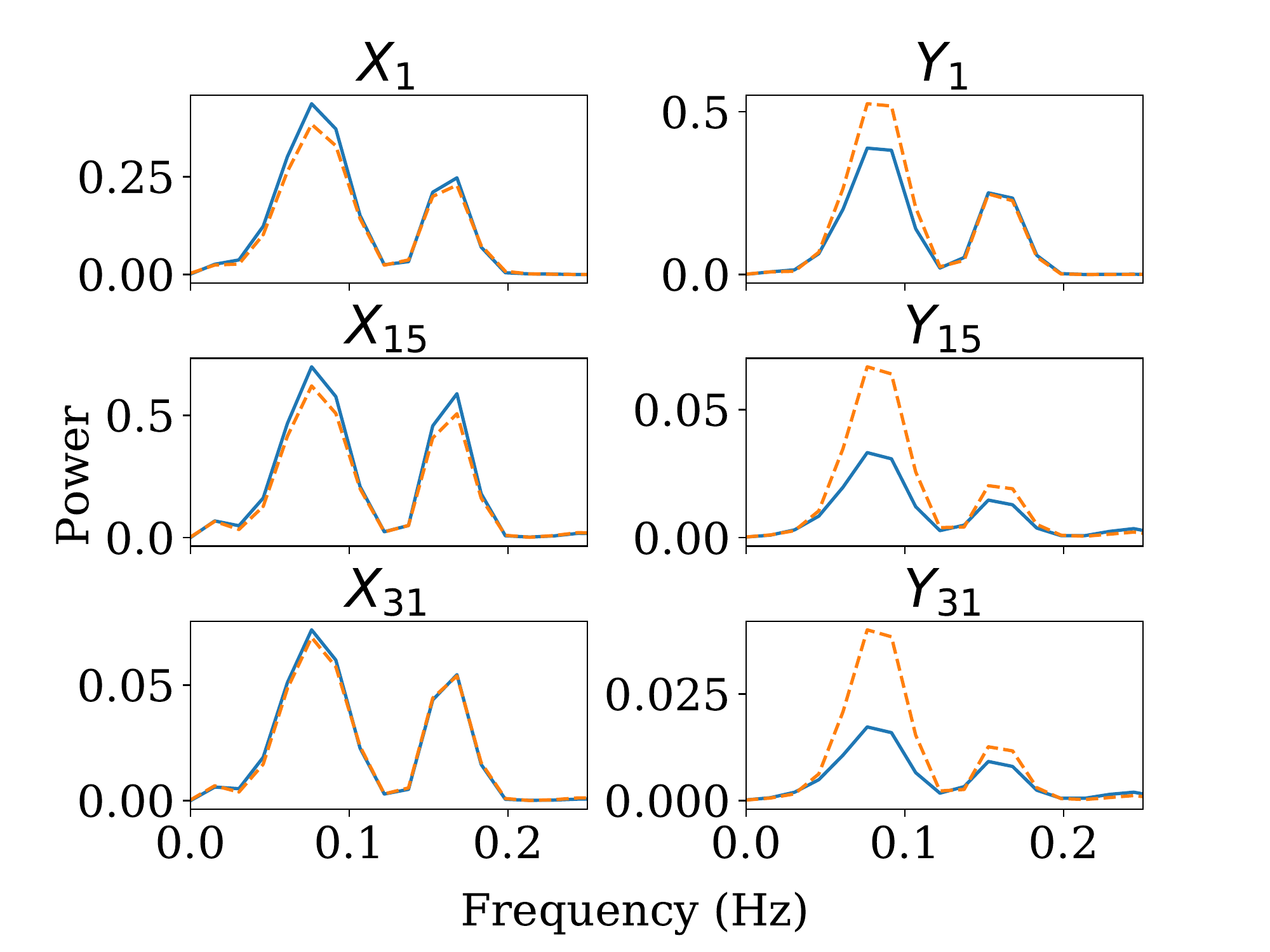}
        \caption{Response spectra}
        \label{fig:ROMtestTransientFFT}
    \end{subfigure}
    \caption{Comparison of the true response (blue line) to the response predicted by the ROM (orange dashed line) in the transient.}
    \label{fig:ROMtestTransientT}
\end{figure}

To test the performance of the trained ROM, we make use of the 12th set of time series, which was not used during training. The trained LSTM network predicts the response for the given input time series, within the latent space. The predicted response in the physical space is then recovered using the Decoder portion of the autoencoder. Figures \ref{fig:ROMss} and \ref{fig:ROMtestTransientT}, compare the response of the ROM with that of the Abaqus simulation for the X and Y displacements at depths of 0, 15 and 30 m. The response is compared both in time and frequency domains.The steady state performance, as demonstrated in Figure \ref{fig:ROMss} of the ROM is good, note that whilst the time series shows only a section of the response for clarity, the spectra is taken from the full response. The normalised mean squared error of the response averaged across all DOFs was found to be 1.87 \% in the steady state (ignoring the first 100 s of simulation). Figure \ref{fig:ROMtestTransient}, in which the first 50 seconds of response and the associated spectra are plotted, shows that once again the performance is significantly worse in the initial transient regime, as can be observed both in the time series and spectra of the response. However, it is considered that the transient regime is not of great concern due to the unrealistic nature of the loading: from zero loading to instantaneous wind and wave loading.

\begin{figure}[t!]
    \centering
    \begin{subfigure}[t]{0.5\textwidth}
        \centering
        \includegraphics[trim={9mm 3mm 20mm 5mm},clip,width=0.95\textwidth]{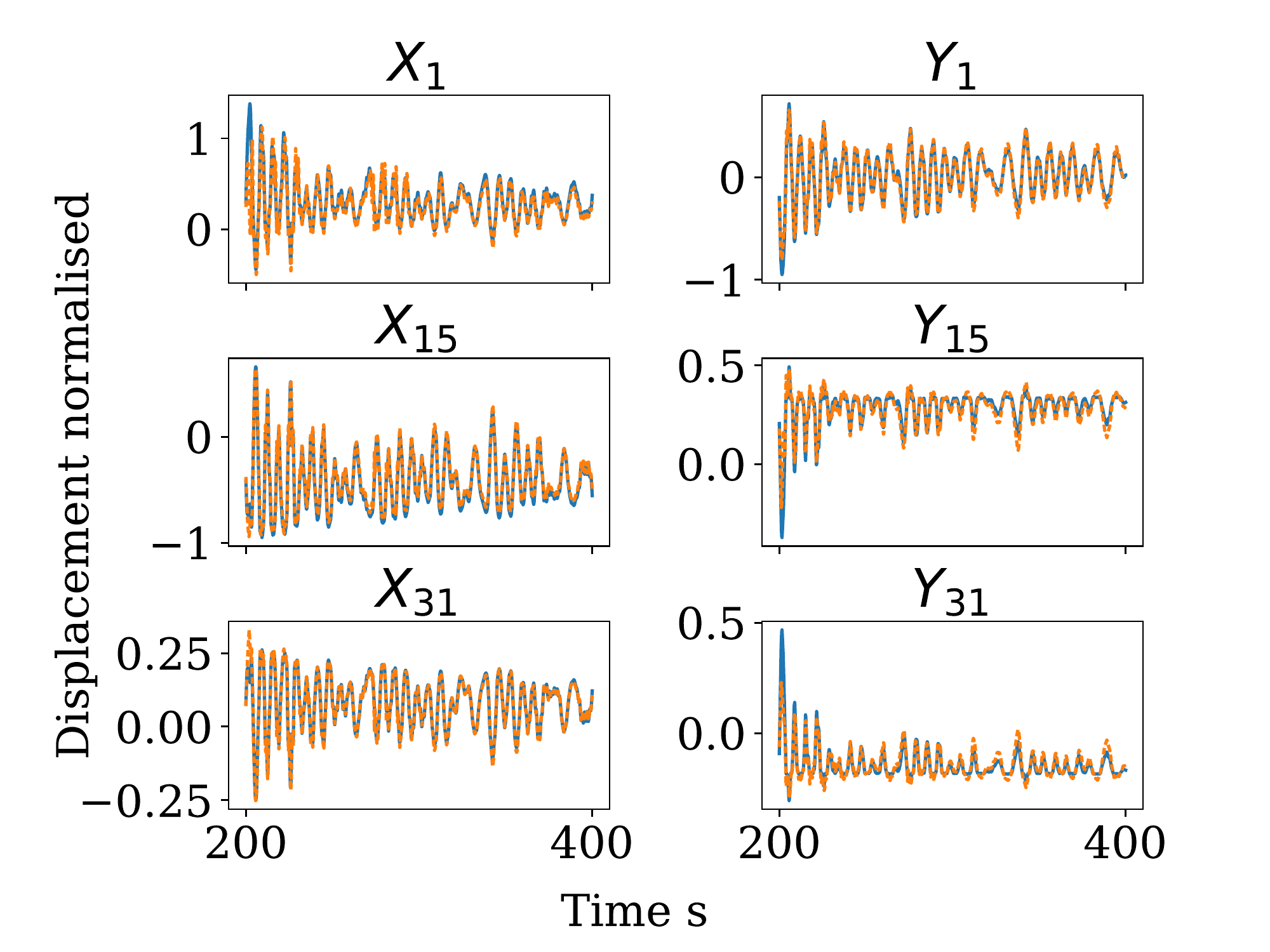}
        \caption{Time series response}
        \label{fig:ROMtestSS}
    \end{subfigure}%
    ~ 
    \begin{subfigure}[t]{0.5\textwidth}
        \centering
        \includegraphics[trim={9mm 3mm 20mm 5mm},clip,width=0.95\textwidth]{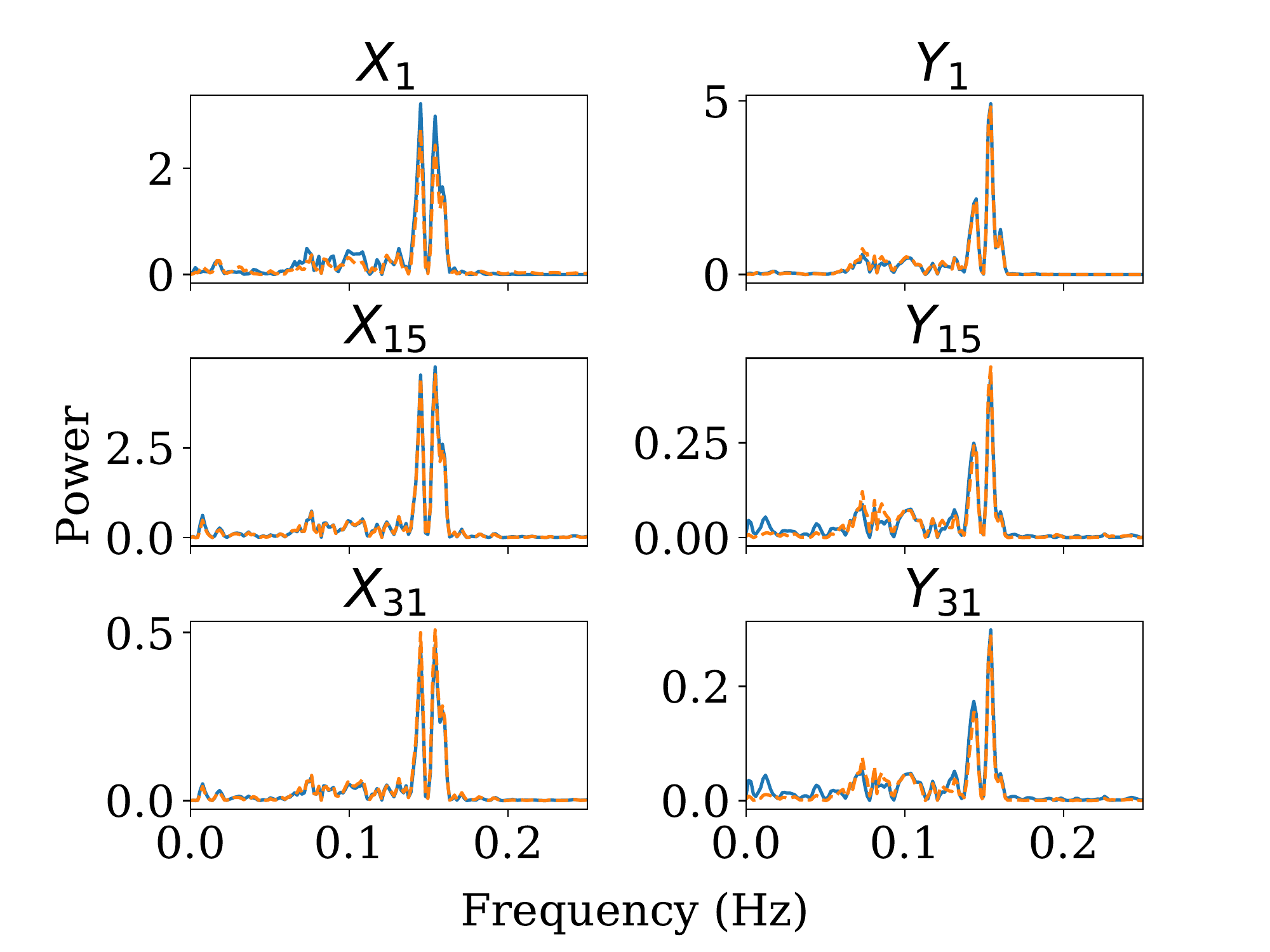}
        \caption{Response spectra}
        \label{fig:ROMtestSSFFT}
    \end{subfigure}
    \caption{Comparison of the true response (blue line) to the response predicted by the ROM (orange dashed line) in steady state.}
    \label{fig:ROMss}
\end{figure}

An important aspect of any ROM is the reduction in computational time compared against the FOM. In this case, a comparison was made between the prediction time of the ROM and Abaqus simulations on the same computing hardware, Intel® Core™ i7-6700 CPU, 3.40GHz 8 core Processor. In both cases a 500 second simulation was carried out and in Table \ref{tab:1} it can be seen that the ROM provides a large benefit in computational time with an over 300 times speed up in computation time.

\begin{table}[h!]
\centering
\begin{tabular}{ll}
\hline
Full Order Abaqus Model    & AE-LSTM ROM \\ \hline
\multicolumn{1}{c}{2272 s} & \multicolumn{1}{c}{6.99 s}  
\end{tabular}
\caption{Computational time of the ROM vs. Abaqus for a 500 s simulation.}
\label{tab:1}
\end{table}

\vspace{12pt}

\noindent \uppercase{\textbf{CONCLUDING REMARKS}} \vspace{12pt}

A newly developed ROM method was demonstrated on a wind turbine monopile system of considerable interest in wind energy systems. The ROM method utilises both Autoencoder and LSTM neural networks to create a data-driven scheme purely relying on input and output data from the system of interest. The method was demonstrated on a monopile structure constrained in a nonlinear soil; represented using nonlinear springs defined from py curves. It was shown that by retaining only 4 latent space variables, a high fidelity ROM could be constructed which accurately predicted steady state response of the structure to a novel forcing input. The method also exhibited significant time improvement when compared to the full scale simulation carried out in Abaqus. Future developments will concentrate on further developing this application-inspired use case to also monitor rotational degrees of freedom and to consider more complex soil models used in the full order simulations.
\newpage
\vspace{12pt}

\noindent \uppercase{\textbf{ACKNOWLEDGEMENTS}} 

\scalerel*{\includegraphics{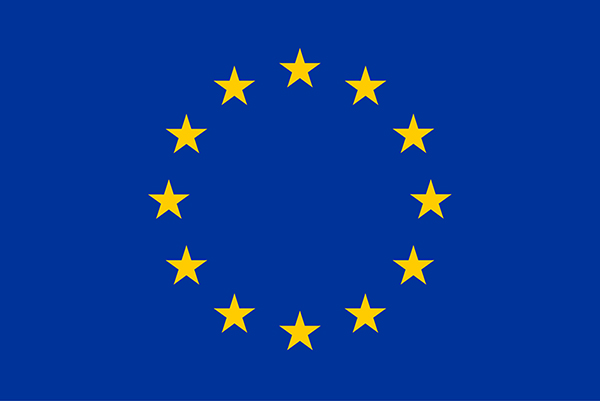}}{2*B} This project has received funding from the European Union’s Horizon 2020 research and innovation programme under the Marie Skłodowska-Curie grant agreement No 764547 and from the EPSRC under grant agreement EP/R004900/1.


\small 

\bibliographystyle{iwshm}
\bibliography{IWSHM_references}
\par\leavevmode
\end{document}